\documentstyle[twocolumn,aps,epsf,psfig,epsfig]{revtex}
\begin{document}
\tighten
\draft

\title{ Very high rotational frequencies and band termination in $^{73}$Br}

\author{
C.\,Plettner$^{1,2}$, H.\,Schnare$^1$, R.\,Schwengner$^1$, 
L.\,K\"aubler$^1$, F. D\"onau$^1$, I.\,Ragnarsson$^3$,
A.V.\,Afanasjev$^{4,5}$, A. Algora$^6$, G.\,de Angelis$^6$, 
A.\,Gadea$^6$, D.R.\,Napoli$^6$, J.\,Eberth$^7$, T.\,Steinhardt$^7$, 
O.\,Thelen$^7$, M.\,Hausmann$^8$, A.\,M\"uller$^8$, A.\,Jungclaus$^8$, 
K.\,P.\,Lieb$^8$, D.G. \,Jenkins$^9$, R.\,Wadsworth$^9$, A.N.\,Wilson$^9$, 
S. Frauendorf$^{10,1}$}

\address{
$^1$ Institut f\"ur Kern- und Hadronenphysik, FZ Rossendorf, 
     01314 Dresden, Germany\\
$^2$ Horia Hulubei National Institute of Physics and Nuclear Engineering,
     Bucharest, P.O. Box MG-6, Romania\\ 
$^3$ Department of Mathematical Physics, Lund Institute of Technology,
     P.O. Box 118, 22100 Lund, Sweden \\
$^4$ Physik-Department der Technischen Universit\"at
     M\"unchen, 85747 Garching, Germany \\
$^5$ Laboratory of Radiation Physics, Institute of Solid State Physics,\\
     University of Latvia, LV 2169 Salaspils, Miera Str. 31, Latvia\\
$^6$ INFN, Laboratori Nazionali di Legnaro, 35020 Legnaro, Italy\\
$^7$ Institut~f\"ur~Kernphysik, Universit\"at~zu~K\"oln,
     50937~K\"oln, Germany\\
$^8$ II. Physikalisches Institut, Universit\"at G\"ottingen,
     37073 G\"ottingen, Germany\\
$^9$ University of York, Physics Department, Heslington,
     York Y01 5DD, United Kingdom \\
$^{10}$ Department of Physics, University of Notre Dame, Notre Dame, IN 46556,
USA 
}
\date{\today}
\maketitle

\begin{abstract}
Rotational bands in $^{73}$Br have been investigated 
up to spins of $I$ = 65/2 using the EUROBALL 
III spectrometer. One of the 
negative-parity bands displays the highest rotational 
frequency $\hbar\omega$ = 1.85 MeV reported to date in nuclei with $A\geq 25$. 
At high frequencies, the experimental ${\cal J}^{(2)}$ dynamic moment 
of inertia for all bands decrease to very low values, 
${\cal J}^{(2)} \leq$ 10 $\hbar^{2}$MeV$^{-1}$. 
The bands are described in the configuration-dependent cranked 
Nilsson-Strutinsky model. The calculations indicate that 
one of the negative-parity bands is observed up to its terminating 
single-particle state at spin 63/2. 
This result establishes the first band termination in the 
$A \approx$ 70 mass region.
\end{abstract}


Exploring nuclei at very high excitation energies and angular momentum 
is of fundamental importance for our 
understanding of many-body systems. Under such extreme 
conditions, one of the most interesting quantum phenomena 
is the termination of rotational bands. This means that a specific 
configuration manifests itself as a collective rotational band at low 
spin values, but gradually loses its collectivity with 
increasing spin and finally terminates at the maximum spin in 
a fully aligned state of single-particle nature 
\cite{Ben83,Ragn95,Afa99}. 
The origin of this phenomenon lies in the limited angular momentum 
content of the fixed configuration. Experimentally, terminating bands 
were first observed in $^{158}$Er \cite{Ragn86}, then 
in the $A \approx 110$ mass region \cite{Ragn95,Jan94} 
and recently in several other mass regions \cite{Afa99}.
In a terminating band, the nuclear shape gradually traces a path 
through the triaxial plane, starting as collective 
(often at near-prolate shape, $\gamma \approx 0^{\circ}$) 
and evolving over many transitions to a non-collective state at 
oblate ($\gamma = +60^{\circ}$), spherical or prolate 
($\gamma = -120^{\circ}$) shape \cite{Ben83}.
In the experiment, this has been most clearly demonstrated
in the $A \approx$ 110 and $A \approx$ 60 mass regions \cite{Wad98,Sve99}.

The A $\approx$ 70-80 mass region displays a large variety of structural 
effects. At low spins, there is a competition between prolate and oblate 
configurations due to the existence of the deformed shell 
gaps at different particle numbers and deformations \cite{Naz85}.
At higher spins, there is a shell gap at near-prolate and near-oblate 
collective shapes at neutron number 38 \cite{Naz85}.
A further interesting observation is that the bands in
$^{81}$Sr \cite{Cri95} show features which are generally associated with 
terminating bands \cite{Afa99,Rag96}, even though they do not appear to 
become fully non-collective.
It is now an interesting question whether bands can be 
observed up to termination in the somewhat lighter
nuclei in the $A = 70$ mass region. 
This is the motivation behind the present study of the 
$^{73}_{35}$Br$_{38}$ nucleus at very high rotational frequencies.

High-spin states in $^{73}$Br were populated in the 
reaction $^{40}$Ca($^{40}$Ca,$\alpha 3p$), using the 185 MeV beam delivered 
by the XTU Tandem accelerator of the Laboratori Nazionali di Legnaro. 
The experiment was performed
using an enriched (99.96 \%) $^{40}$Ca target with a thickness of 
0.9 mg/cm$^2$. $\gamma$-rays were registered with
15 Cluster \cite{Ebe96} and 26 Clover \cite{Bec94} detectors of 
the EUROBALL III array. Charged particles were detected with the 
Italian SIlicon Sphere (ISIS) consisting of 40 Si $\Delta E-E$ telescopes 
\cite{Far97}. At forward angles, 15 segmented detector units filled with
BC501A liquid scintillator were mounted to detect neutrons \cite{Ske99}.

\begin{figure}
\epsfig{file=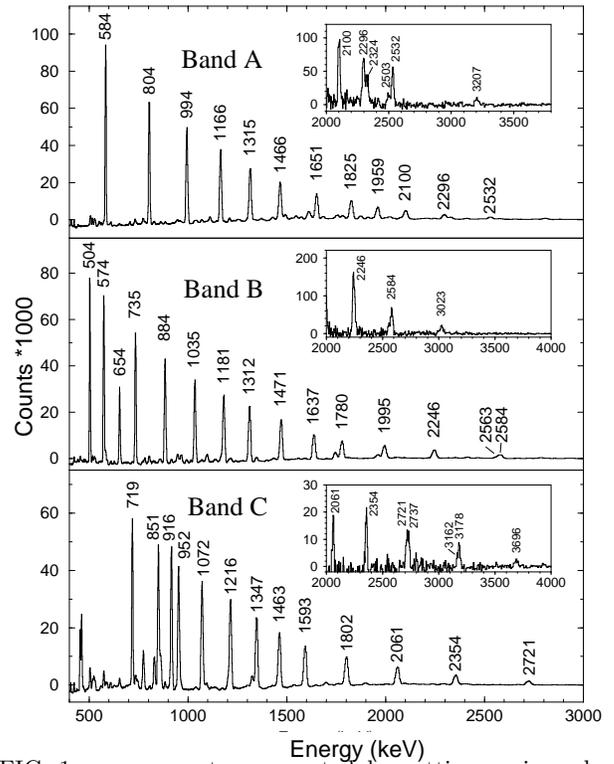,width=8cm}
\caption{$\gamma$-ray spectra generated by setting various double-gates 
in the cube. The insets show the higher energy part of the corresponding 
spectra.}
\label{fig:cubespec}
\end{figure}

\begin{figure}
\psfig{file=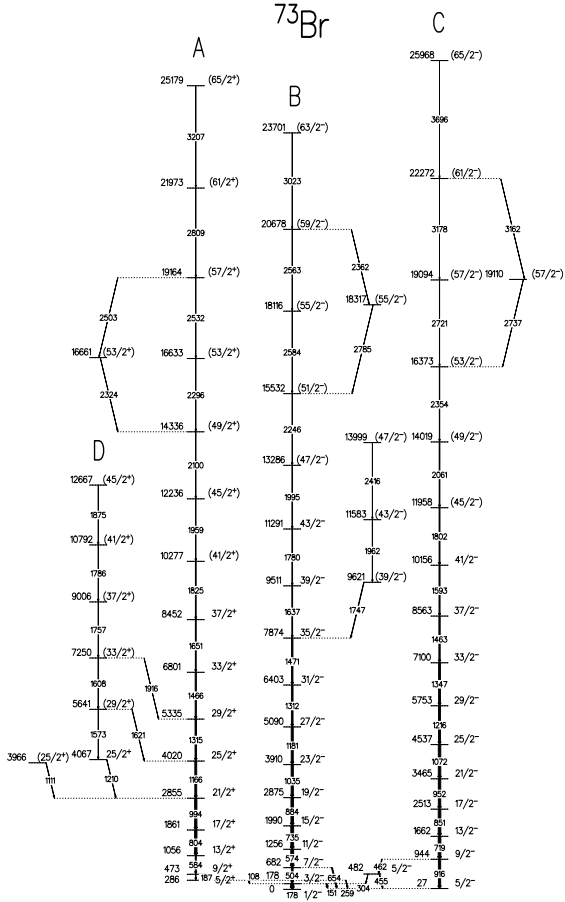,width=9cm,height=13cm}
\caption{Level scheme of $^{73}$Br deduced from the present experiment.
 For the band-head discussion see Ref. \protect\cite{Hee90}.}
\label{fig:brlschem}
\end{figure}

A total number of 2$\times$10$^{9}$  $\gamma\gamma\gamma$ events 
were recorded. The $\alpha3p$ exit channel leading to
$^{73}$Br is predicted to 3.4\% of the total cross section, according
to PACE calculations \cite{Gav80}. 
$\gamma\gamma$-particle coincidences and 
$\gamma\gamma\gamma$ coincidences were sorted into 
two-dimensional $E_{\gamma}-E_{\gamma}$ matrices and 
three-dimensional $E_{\gamma}-E_{\gamma}-E_{\gamma}$ cubes.
Examples of doubly-gated coincidence spectra are shown in Fig. 
\ref{fig:cubespec}.
These coincidence data were analysed
using the Radware package \cite{Rad95}. The resulting level scheme of
the $^{73}$Br nucleus is presented in Fig. \ref{fig:brlschem}.
The sequences A, B and C have been known from previous studies 
\cite{Hee87,Hee90} up to the states of spins and parities $I^{\pi}$ = 
(45/2$^{+}$), (47/2$^{-}$) and (49/2$^{-}$), respectively. In this work, 
these rotational bands were extended up
to states with $I^{\pi}$ = (65/2$^{+}$), (63/2$^{-}$) and 
(65/2$^{-}$), respectively, at excitation energies $E$ $\approx$ 26 MeV. 
Moreover, a new sequence D was established, which feeds the 2855 keV,
4020 keV and 5335 keV states of band A.

\begin{figure}
\epsfig{file=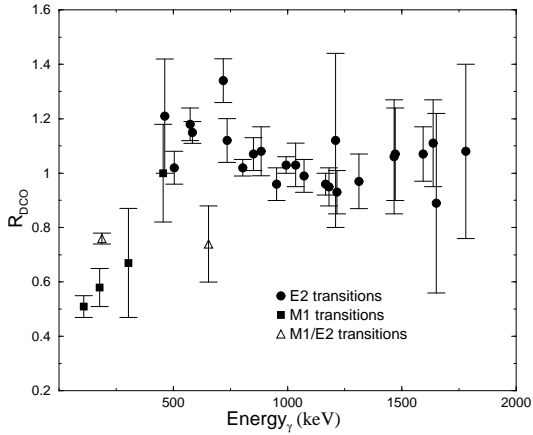,angle=-90,width=7cm}
\caption{DCO ratios for transitions in $^{73}$Br. The gate is set on an 
E2 transition.}
\label{fig:dco}
\end{figure}

An analysis of the directional correlations from oriented states (DCO) 
was performed to assign spins to the newly observed 
states. 
The coincidence data were added up for the 35 most backward-angle detectors, 
located at an average angle of 156$^{\circ}$ (149$^{\circ}$, 155$^{\circ}$, 
157$^{\circ}$, 163$^{\circ}$) to the beam and the 108 detectors 
near 90$^{\circ}$ (72$^{\circ}$, 81$^{\circ}$, 99$^{\circ}$, 107$^{\circ}$).
A 156$^{\circ}$ versus 90$^{\circ}$ $\gamma\gamma$ 
matrix was created in coincidence with $\alpha$ particles. 
From this matrix we extracted DCO-ratios defined as:
\begin{eqnarray}
R_{\rm DCO}^{exp} = \frac{I_{156^{\circ}}^{\gamma_2}(Gate^{\gamma_1}_{90^
{\circ}})}{I_{90^{\circ}}^{\gamma_2}(Gate^{\gamma_1}_{156^{\circ}})}
\end{eqnarray}
where $I_{\rm 156^{\circ}}^{\gamma_2}(Gate^{\gamma_1}_{90^{\circ}})$ 
denotes the efficiency-corrected intensity of the $\gamma_2$ transition 
observed at 156$^{\circ}$ when gating on $\gamma_{1}$ at 90$^{\circ}$. 
According to the calculations for the DCO-ratios by Krane et 
al. \cite{Kra73}, values of about 0.5 and 1 are expected for stretched 
and pure transitions of multipole order 1 and 2, respectively, if the gate
is set on an $E$2 transition. 
The DCO-ratios of transitions in $^{73}$Br are plotted in Fig. \ref{fig:dco}. 
There are two clearly separated groups of transitions
around DCO-ratios of 0.6 and 1, which are assigned as
dipole and quadrupole transitions, respectively.
The DCO-ratio very close to 1 for the 455 keV M1 transition is related
to the spin difference $\Delta I = 0$
between the initial and final states (see Fig. \ref{fig:brlschem}) .
On the basis of the DCO-ratios we confirmed the character of all previously 
known transitions. We furthermore firmly derived an E2 character for the 
1651 keV $\gamma$-ray in band A, for the 1471 keV, 1637 keV, 1780 keV 
$\gamma$-rays in band B, for the 462 keV and 1593 keV $\gamma$-rays in band C, 
and for the 1210 keV $\gamma$-ray depopulating the lowest 
observed state in band D. Hence, band D has the same signature and parity 
as band A.

The rotational bands in $^{73}$Br display very high $\gamma$-ray energies
(see Fig. \ref{fig:brlschem}). For example the $\gamma$-ray energy on the 
top of sequence C is 3696 keV. This corresponds to a rotational 
frequency of $\hbar\omega$ = $E_{\rm \gamma}$/2 = 1.85 MeV. This is the 
highest frequency ever observed in a rotational 
cascade in nuclei with $A \geq 25$. 
For comparison, the highest rotational frequencies reported so far 
are $\hbar\omega$ = 1.82 MeV in the $^{60}$Zn nucleus \cite{Sve99}
and $\hbar\omega$ = 1.4 MeV in the $^{109}$Sb nucleus \cite{Sch96}.

\begin{figure}
\epsfig{file=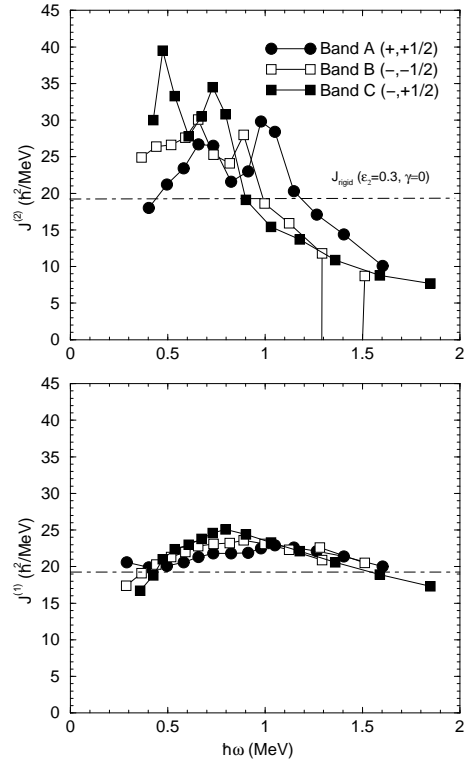,width=6cm}
\caption{Observed dynamical moment of inertia, ${\cal J}^{(2)}$ and
kinematical moment of inertia ${\cal J}^{(1)}$, as a function
of frequency for the bands A, B and C. The horizontal dot-dashed line 
indicates the rigid-body value for a deformation of $\varepsilon_2$ = 0.3 and 
$\gamma$ = 0.}
\label{fig:j2}
\end{figure}

A collective parameter which is very sensitive to the changes in the
nuclear structure is the dynamic moment of inertia, 
${\cal J}^{(2)} = (dE_{\gamma}/dI)^{-1}$. In Fig. \ref{fig:cubespec} 
a gradual increase in the $\gamma$-ray energy spacings as the 
$\gamma$-ray energies increase within the bands can be clearly seen. 
This implies a corresponding decrease of the dynamic moments of inertia 
${\cal J}^{(2)}$, which are presented in Fig. \ref{fig:j2}-top. 
For $\hbar\omega \leq $ 1 MeV, the irregularities of 
${\cal J}^{(2)}$ are signaling band crossings. 
For each band there are mainly two irregularities in ${\cal J}^{(2)}$
which could be caused by proton and neutron $g_{9/2}$ alignments.
In contrast, for frequencies higher than $\approx$ 1 MeV, 
the dynamic moments of inertia of the bands converge 
and decrease to approximately 40 \% of the value of a rigid rotor.
This smooth down-sloping of the dynamic moment of inertia may indicate 
that, starting at $\hbar\omega \approx 1.0$ MeV, the configurations of 
bands A and C do not change up to the highest rotational frequency, 
and that the rotational band is gradually losing 
its collectivity. For band B a negative spike in ${\cal J}^{(2)}$  
occurrs at $\hbar\omega$ = 1.28 MeV and is caused by the decrease of the 
transition energy which indicates a band crossing.
This behavior is reminiscent of $^{158}$Er where a band terminating
at $I = 46^+$ crosses the more collective yrast band at
$I \approx 40$ \cite{Ragn86}. The bottom graph of Fig. \ref{fig:j2} 
shows the kinematical moment of inertia ${\cal J}^{(1)}$ = $I/\omega$. 
The latter stays close to the rigid-body value. 
For high rotational frequencies the relation ${\cal J}^{(1)} \gg {\cal J}
^{(2)}$ indicates that pairing correlations play no important role.    

In order to assign configurations to bands A, B and C 
we have used the configuration-dependent Cranked 
Nilsson-Strutinsky (CNS) approach \cite{Ben85,Afa95} 
based on the cranked Nilsson potential. Since we are 
interested in the high-spin properties, the pairing 
correlations have been neglected in the calculations.
The calculations minimize the total energy of a specific 
configuration for a given
spin with respect to the deformation parameters 
($\varepsilon_{2},\varepsilon_{4},\gamma$). Thus, the total energy 
and the shape trajectory (the evolution of the minimum of the total energy
in the ($\varepsilon_{2},\gamma$) plane as a function of spin) are obtained 
for each configuration. The configurations of interest can be described 
by excitations within the $N = 3$ $p_{3/2}$, $f_{5/2}$, $p_{1/2}$  
and  the $N = 4$ $g_{9/2}$ orbitals. Thus, there are no holes 
in the $f_{7/2}$ subshell and no particles 
in the orbitals above the spherical shell gap at 50.
In the Ref. \cite{Naz85} it was shown that 
at the deformation of $\varepsilon_2$ = 0.35 the lowest neutron 
intruder orbital from the $h_{11/2}$ subshells comes
down and crosses the Fermi surface at $\hbar\omega$ = 1.4 MeV.
However, the CNS calculations including one neutron in the
$h_{11/2}$ orbital result in bands which behave very differently
from the observed bands. 

Furthermore, the highest $N = 3$ $p_{1/2}$ orbital
will not become occupied in the low-lying configurations of
$^{73}$Br; therefore we can omit $p_{1/2}$ in the labeling.
Note also that all orbitals are treated on the same footing in 
the cranking calculations, which means that, for example, 
the polarization of the core is taken care of. 

In the following, the configurations will be specified with respect to a 
$^{56}_{28}$Ni$_{28}$ core as having 7 active protons and 10 active 
neutrons. They will be labeled by the shorthand 
notation  [$p_1p_2,n_1n_2$], where $p_1(n_1)$ 
stands for the number of protons (neutrons) in the ($p_{3/2},f_{5/2}$)
orbitals and $p_2(n_2)$ stands for the number of protons 
(neutrons) in $g_{9/2}$ orbitals. In addition, the sign 
of the signature $\alpha$ of the last occupied orbital (given as 
a subscript) is used if the number of occupied orbitals
in the specific group is odd.

Considering that the bands A, B and C extend to high spins of $I = (65/2)$
or $(63/2)$, we will first outline the possible proton and neutron 
configurations using their maximum spins $I_{\rm max}^{p,n}$ as criteria. 
The maximum spin is defined from the distribution of particles 
and holes over the $j$-shells at low spin.
Including 2 protons in the $g_{9/2}$ orbital, there are 5 
$(f_{5/2}p_{3/2})$ protons, which lead to a maximum proton spin of   
$I_{\rm max}^{p}= 27/2$ or 29/2 for the proton
subsystem depending on signature. The maximum proton spin becomes 33/2 for 
proton configurations with 3 $g_{9/2}$ protons if the last 
proton occupies the favoured ($\alpha = +1/2$) $g_{9/2}$ signature. 
Including 3 neutrons in the $g_{9/2}$ orbital, the maximum neutron spin 
$I_{\rm max}^{n}$ is 15 or 16 depending on the signature for the 7 
$(f_{5/2} p_{3/2})$ neutrons. Adding one more neutron to the $g_{9/2}$ 
orbital, the maximum neutron spin is $I_{\rm max}^{n} = 18$.
The occupation of additional $g_{9/2}$ orbitals adds only 
marginally to the maximum spin, but costs a lot of energy. 
Thus, it is reasonable to expect that such configurations 
will be considerably above the yrast line in the spin range 
of interest. There are five out of all possible combinations, where the
proton and the neutron spins add to a maximum total spin
of $I_{\rm max} \geq 63/2$.
The calculations show that one of these configurations, 
[43$_{+}$,7$_{-}$3$_{+}$], 
does not build any `collective band' for spin values $I \approx 30$. 
Therefore, only four configurations,
namely [43$_{+}$,64], [5$_{+}$2,64], [5$_{-}$2,64]
and [43$_{+}$,7$_{+}$3$_{+}$], are left as possible candidates
for the observed bands. 

\begin{figure}
\epsfig{file=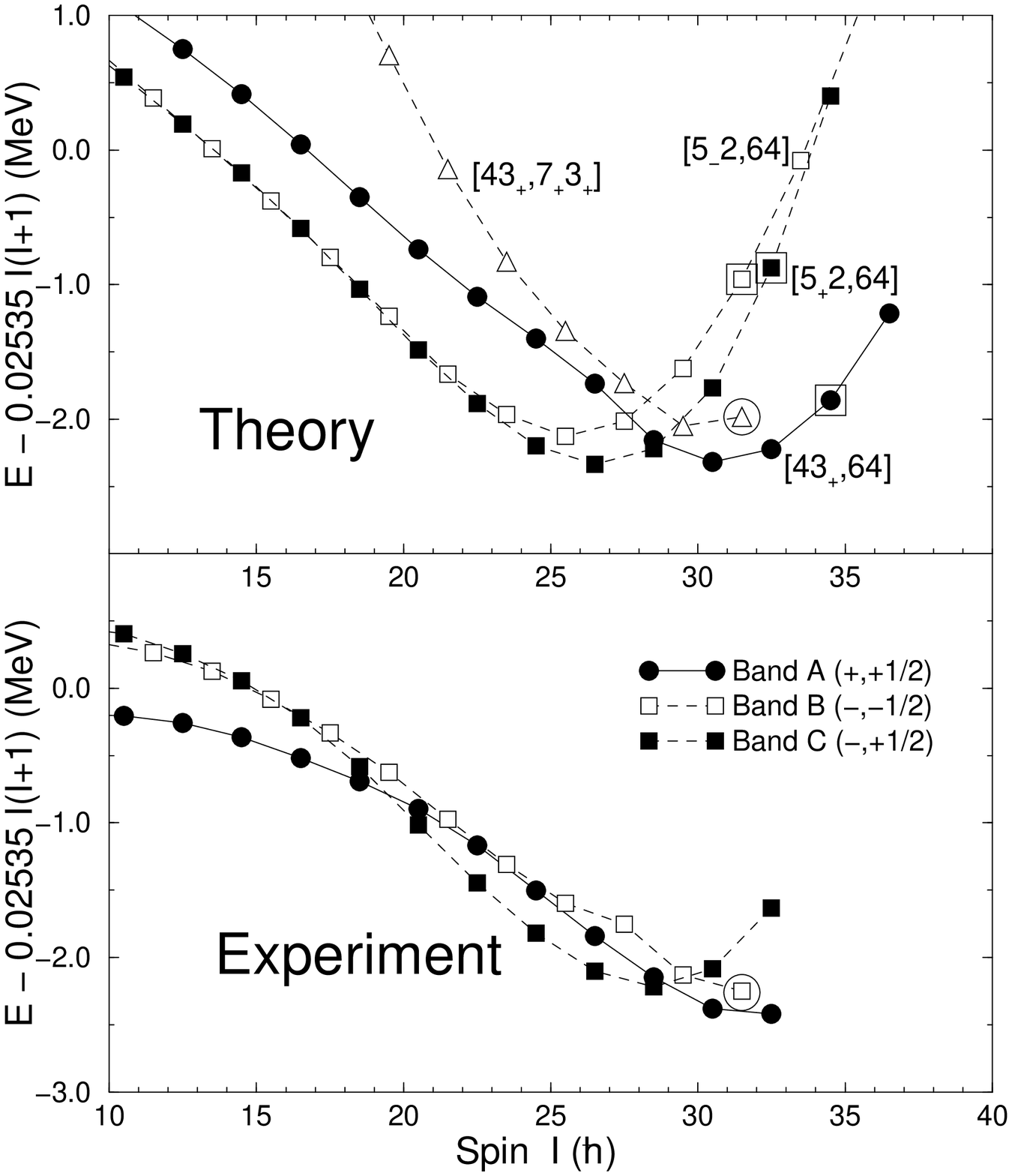,width=7cm}
\caption{Comparison between the experimental (lower panel) and calculated
(upper panel) energies relative to a rigid rotor reference.
The parameters for $A$ = 80 nuclei from \protect\cite{Gal86} 
were used in the CNS calculations. The calculated energies 
are normalized so that the energy of the [5$_{+}$2,64] 
configuration for spin $I$ = 57/2 fits the observed energy of 
band C. Terminating states are indicated by large open circles and 
collective states of `maximum spin', $I_{\rm max}$, by large open squares.
In the spin range $I$ = 41/2-57/2, the configurations
[5$_{+}$2,7$_{-}$3$_{+}$] and [5$_{-}$2,7$_{+}$3$_{+}$], 
terminating in non-collective oblate states at $I$ = 45/2 and $I$ = 57/2,
 are calculated to be somewhat lower than the configurations shown in 
the figure. They are nevertheless not included, because their maximum 
spins are below the maximum spins observed in the experiment.}
\label{fig:exptheo}
\end{figure}

In Fig. \ref{fig:exptheo} the calculated energies relative
to a rigid rotor reference $(E-E_{\rm RR})$ of the above mentioned 
configurations are compared with the experimental data. The CNS 
calculations indicate that these configurations are indeed the 
lowest-lying ones capable of building angular momentum up to the values 
observed. Since pairing correlations have been neglected, the calculated 
energies are expected to be realistic only at high spin of $I \geq 15$.
In the calculations, the configurations are following a parabola-like 
energy curve. As seen in Fig. \ref{fig:exptheo}, the [43$_{+}$,64] 
configuration has its minimum around spin 61/2, while band A seems to
approach the minimum at 65/2. Amongst the considered configurations, 
the [5$_{+}$2,64] configuration is calculated to be lowest up to spin 
$I = 57/2$, which reproduces the experimental situation with 
band C. Moreover, the minimum in $(E-E_{\rm RR})$
occurs at similar spin values, 57/2 in band C and 
at 53/2 in the [5$_{+}$2,64] configuration. Based on this comparison of 
theoretical and experimental energies we assign the [43$_{+}$,64] and 
[5$_{+}$2,64] configurations to bands A and C, respectively.   

\begin{figure}
\epsfig{file=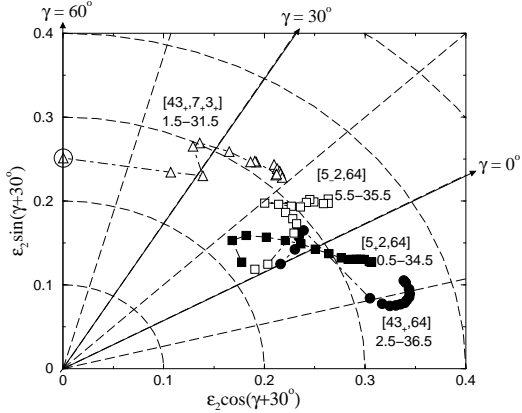,width=7cm}
\caption{Calculated shape trajectories in the ($\varepsilon_2,\gamma$) plane
for the configurations assigned to the bands A, B and C.
 The range of the angular momentum is given next to the configuration.
The trajectories are not regular at low spin due to the strong competition
and coexistence between different minima in the potential
energy surfaces corresponding to positive or negative $\gamma$ values.} 
\label{fig:shapes}
\end{figure}

Band B is observed up to $I$ = (63/2).
The calculations suggest that the [5$_{-}$2,64] 
configuration can be assigned to this band at low and 
medium spin. It is the signature partner of the [5$_{+}$2,64] 
configuration assigned to band C. These configurations 
reproduce well the slope of experimental $(E-E_{\rm RR})$ curves 
for bands B and C, but underestimate the signature splitting between them.
The [5$_{-}$2,64] configuration is 
crossed at spin 57/2 by the [43$_{+}$,7$_{+}$3$_{+}$] configuration
(see Fig. \ref{fig:exptheo}). The predicted crossing is indeed observed 
in band B at spin (55/2$^{-}$) (Fig. \ref{fig:brlschem}). Thus, the latter 
configuration can be related to band B above the band crossing. 
In the calculations, the band built on the [43$_{+}$,7$_{+}$3$_{+}$] 
configuration terminates at the maximum spin of the particle configuration,
i.e. at 63/2. This can be seen in the shape trajectories, which are presented 
in Fig. \ref{fig:shapes}. Whereas the trajectories related to bands A and C 
include collective near-prolate deformations over the whole spin range, 
the configuration assigned to band B after the band crossing at 55/2 
(empty triangles) undergoes a shape change from a collective 
$\gamma \approx +30^\circ$ shape to a non-collective oblate 
$\gamma = +60^\circ$ shape between spins 55/2 and 63/2. Thus, the band built 
on this configuration terminates at the maximum spin $I_{\rm max} = 63/2$.
Since this coincides with the maximum spin observed in band B, we have 
observed this band up to its termination.
Our interpretation of band B requires a relatively strong interaction 
between two configurations which differ in their occupation of the 
$j$-shells for both protons and neutrons. This might appear unlikely
but one could also note that a high-spin crossing in
a terminating band in $^{108}$Sn has been interpreted \cite{Wad96}
as built from configurations which differ in an analogous
manner.

In band A a branching at spin $I$ = (53/2) is experimentally seen  
 (see Fig. \ref{fig:brlschem}). The configuration 
assigned to this band shows a shape change from $\gamma \approx -15^{\circ}$ 
to $\gamma \approx +5^{\circ}$ (see Fig. \ref{fig:shapes}) between
$I$ = 53/2 and 57/2.
This suggests that in each minimum, there is a smooth configuration
and the calculated yrast configuration jumps from one minimum to the other 
with increasing spin. 
Thus, the two states at (53/2) might belong to
these different minima. A similar 
branching is also observed for the band C, but appears even more
difficult to 
describe in the calculations even though the irregularities
in the shape trajectories (Fig. \ref{fig:shapes}) suggest the possibility of
competitions between different minima in all configurations
shown in Figs. \ref{fig:exptheo} and \ref{fig:shapes}. 
However, these branchings remain an interesting experimental result
and a challenge to theory.

In summary, we observed in the $^{73}$Br nucleus the highest
rotational frequency in a cascade in nuclei with $A \ge 25$. 
Out of the four observed rotational bands could be followed up to its 
termination. Moreover, we give arguments on the particle structure 
of this band. 
According to the calculations the other two high-spin bands are not 
terminating; instead they stay collective beyond the maximum spin 
defined from the distribution of particles and holes 
over the $j$-shells at low spin.
These results establish the phenomenon of 
band termination in the $A\approx 70$ mass region for the first time. 
The predicted decrease in the quadrupole moment along the terminating
band should be observable via Doppler shift lifetime measurements.
Further studies will be important in order to obtain greater insight into 
the systematics of the physical observables in connection with band 
termination in this mass region. 

This work was supported by the German Ministry of Education and Research 
under contracts 06 DR 827, 06 OK 862, 06 GOE 851, by the 
Swedish Natural Science Research Council, by the UK EPSRC, and by the 
European Union within the TMR project. A.V. Afanasjev acknowledges the 
support by the Alexander von Humboldt Foundation.

\end{document}